\documentclass[%
 aip,
 jmp,%
 amsmath,amssymb,
  reprint,%
 author-numerical,%
]{revtex4-1}

\usepackage{graphicx}
\usepackage{dcolumn}
\usepackage{bm}

\begin{document}

\preprint{}

\title[Sample title]{Synchronizing spatio-temporal chaos
with imperfect models: a stochastic surface growth picture}

\author{Diego Paz\'o}
\author{Juan M. L\'opez}
\affiliation{Instituto de F{\'\i}sica de Cantabria (IFCA),
CSIC-Universidad de Cantabria, 39005 Santander, Spain}

\author{Rafael Gallego}
\affiliation{Departamento de Matem\'aticas, Universidad de Oviedo, Campus de
Viesques, 33203 Gij\'on, Spain}

\author{Miguel A. Rodr{\'\i}guez}
\affiliation{Instituto de F{\'\i}sica de Cantabria (IFCA),
CSIC-Universidad de Cantabria, 39005 Santander, Spain}

%

\date{\today}

\begin{abstract}
We study the synchronization of two spatially extended dynamical systems where
the models have imperfections. We show that the synchronization error across
space can be visualized as a rough surface governed by the Kardar-Parisi-Zhang
equation with both upper and lower bounding walls corresponding to
nonlinearities and model discrepancies, respectively. Two types of
model imperfections are considered: parameter mismatch and unresolved fast scales, finding in
both cases the same qualitative results. The consistency between different
setups and systems indicates the results are generic for a wide family of
spatially extended systems.
\end{abstract}

 \keywords{Chaotic synchronization, Spatio-temporal chaos, bounded KPZ equation} 
\maketitle

\begin{quotation}
Identical chaotic systems are able to perfectly synchronize when coupled. This
phenomenon is very attractive from a theoretical perspective and has a
tremendous potential for technological applications, like for instance, secure
optical communications. Chaotic synchronization of spatially extended dynamical
systems plays also a fundamental role in forecasting applications and
observation data assimilation in geoscience. Unfortunately, dynamical systems
are far from identical in practical applications and synchronization of high
dimensional systems in the real world becomes severely hampered by
imperfections like parameter mismatches and unresolved scales. Understanding
the bounds to synchronization for imperfect systems, the effect of finite
non-small parameter deviations, and limited resolution on the statistics of the
synchronization error is essential for real applications. The theoretical
challenge is to describe synchronization of imperfect models under the umbrella
of a generic stochastic theory that describes the most outstanding features in a
model--independent fashion.
\end{quotation}

\section{Introduction}

Synchronization phenomena between two identical chaotic systems have been
exhaustively investigated since the 1990s~\cite{pecora97,PRK01,boccaletti02}. It
was soon realized that unavoidable differences between two interacting systems
$\mathbf{u}$ and $\mathbf{v}$ makes generically impossible to observe complete
synchronization, {\it i.e.} $\mathbf{w}(t)\equiv\mathbf{v}(t)-\mathbf{u}(t)=0$,
in real systems. This led to the concept of `attractor
bubbling'~\cite{ashwin94}, which is successfully explained in terms of certain
unstable periodic orbits~\cite{heagy95,venkataramani96}. However, in many
practical situations the systems under investigation are spatially extended (or
high-dimensional) and strongly chaotic, rendering that theoretical framework
clearly inadequate. A prominent example of the latter is the coupling of lasers
with optical feedback, customarily used in encrypted communications. The problem
of synchronization of two spatially extended nonidentical dynamical systems has
been considered before~\cite{boccaletti99,duane01,ginelli09} (see
also~\cite{revuelta,cohen08,huang09} for examples of time-delayed systems). Most
of these previous works made essentially phenomenological explorations, except
for the theoretical work of Ginelli {\it et al.}~\cite{ginelli09} where a
certain scaling relation was found for an infinitesimal parameter mismatch.

The purpose of the present study is to provide a qualitative picture of partial
synchronization of spatially extended nonidentical systems for finite {\em
non-small} parameter mismatch and/or significant model imperfections. Our aim is
to obtain a stochastic field-theoretical description of synchronization for
dynamical systems with significant dissimilarities and explore to what extent
this field theory is valid for different types of model discrepancies. Such a
theoretical description will enhance our understanding on how partial
synchronization is achieved, to what degree, and what are the limiting factors
to synchronize nonidentical systems in real-world applications.

Our approach builds upon the seminal work of Ahlers and Pikovsky~\cite{ahlers02}
that made a connection between the synchronization error of two coupled
identical spatially extended systems and the dynamics of rough surfaces
described by the `bounded Kardar-Parisi-Zhang' (bKPZ) equation~\cite{tu97}.
Interestingly, the bKPZ class also characterizes the synchronization error
between two (identical) time-delayed systems~\cite{szendro05}. In the presence
of dissimilarities of the models, our framework permits to qualitatively
understand the dependence of the synchronization error statistics on the
coupling strength.

Although not crucial for our main conclusions, in this work we consider a
master-slave (also called drive-response or emitter-receiver) configuration
where system $\mathbf{u}$ (the master) forces $\mathbf{v}$ (the slave). This
synchronization setup, so-called unidirectional coupling, is used, for instance,
in encrypted communications with chaos~\cite{roy98,argyris05}. Remarkably,
unidirectional coupling is also relevant in certain data assimilation techniques
that are widely used in geoscience. Indeed, the technique termed {\em
nudging}~\cite{hoke76} is a classical data-assimilation method in which the
unavoidably imperfect model is coupled to the `truth' or `reality' using
(imperfect and sparse) observations. By means of unidirectional coupling the
model ``assimilates'' the reality, and the model's variables serve as an
estimator of the state of the truth (say the atmosphere or the ocean) from
observations~\cite{Kalnay}. Unsurprisingly, the relationship between
synchronization and `nudging' has been highlighted and studied in the
past~\cite{yang06,duane06,szendro_jgr,abarbanel10}.

The paper is organized as follows. In Sec.~\ref{models} we present the models we
have used in our numerical simulations, and in Sec.~\ref{review} we review the
relation between the bKPZ equation and complete synchronization of
spatio-temporal chaos. In Secs.~\ref{mismatch} and~\ref{unresolved} we present
our numerical results for two typical situations: the case of a slave with
significant parameter mismatch with respect to the master and, on the other
hand, a situation where the slave is lacking the variables corresponding to the
small length scale dynamics. The latter is specially relevant in some
data assimilation applications in geoscience, where the models that are
``coupled'' to reality only describe the atmosphere or ocean state above a
certain scale~\cite{yang06,duane06,szendro_jgr,abarbanel10}. Finally, in
Sec.~\ref{concl} we interpret our results by the light of a surface picture
of the synchronization error dynamics and summarize them.

\section{Models}
\label{models}

Let us first start summarizing the type of master-slave configurations we will
consider in this paper. Typically, our results will apply to systems that
fulfill the following properties:

\begin{itemize}
\item [{\it (i)}] The master and the slave are spatially extended systems with
spatio-temporally chaotic dynamics. They can be either continuous or discrete in
both space and time.

\item [{\it (ii)}] The systems are assumed to be extended in one lateral spatial
dimension (though not fundamental differences should arise in larger
dimensions). Note that time-delayed systems belong to this category in virtue of
a coordinate transformation, see Sec.~\ref{delay}.

\item [{\it (iii)}] The equations governing the master and the (uncoupled) slave
systems are nonidentical, due to either a parameter mismatch or the existence of
unresolved degrees of freedom.

\item [{\it (iv)}] The coupling between both systems is dense (ideally at all
points). Note that, from an experimental point of view, this is more easily
achieved for coupled time-delayed systems than for real
spatially extended systems. Actually, the premise of a dense constant coupling
is generally assumed in any study of identical systems synchronization that we
are aware of.

\end{itemize}

The system sizes we use in our simulations are large enough to allow the
systems to exhibit spatially extended chaos. As we shall see, the
synchronization threshold is not a critical point in the case of nonsmall
parameter mismatch and, therefore, there are not finite-size corrections to
scaling to take care of or other subtleties associated with critical point
phenomena. This allows us to obtain generic properties with relatively modest
system sizes, as compared with those needed in studies of synchronization of
identical systems at the critical point. 

In the remainder of this section we introduce the models used in the numerical
simulations presented in this paper.

\subsection{Spatially-extended systems with parameter mismatch}

\subsubsection{Coupled-map lattice}

Coupled-map lattices (CMLs) are particularly efficient for numerical simulations of
spatio-temporal chaos. In a master-slave configuration we have 
\begin{equation}
u(x,t+1)=(1+\varepsilon {\cal D}) f_m(u(x,t)), \label{cmlm}
\end{equation}
for the master, and
\begin{eqnarray}
v(x,t+1)&=&(1+\varepsilon {\cal D}) [\gamma f_m(u(x,t)) + \nonumber \\
&+& (1-\gamma)
f_s(v(x,t))], 
\label{cmls} 
\end{eqnarray}
for the slave. $\cal D$ is the discrete Laplacian:
$$
{\cal D} G(x)=G(x-1)-2G(x)+G(x+1),
$$
and periodic boundary conditions are assumed in the spatial coordinate
$x=1,\ldots,L$. We adopt the fixed value $\varepsilon=1/3$ for the diffusion
coefficient. We vary the coupling parameter $\gamma$ that controls the input
strength of the master on the slave. In this work we choose the system size to
be $L=4096$, which is
large enough to capture
generic phenomena in extended systems.
The local
dynamics is chosen to obey the logistic map $f_{m,s}(z)=\mu_{m,s}-z^2$, with
$\mu_m=1.9$ (for the master) and $\mu_s=\mu_m+\Delta\mu$ (for the slave) taking
values corresponding to a chaotic parameter region of the map. We study the effect
of a finite and non-small parameter mismatch $\Delta\mu$ on the synchronization
error.

\subsubsection{Lorenz-96 model}

The Lorenz-96 model~\cite{lorenz96,lorenz98} was originally introduced as a toy
model of the atmosphere consisting of $L$ nodes each representing a scalar
variable at one site on a latitude circle. The model is, moreover, consistent
with a discretization of the damped Burgers-Hopf equation under
forcing~\cite{Majda}: $\partial_tu-u \partial_x u =- \gamma \,u+F$.
Unidirectional coupling between two Lorenz-96 models yields the following system
of ordinary differential equations:
\begin{eqnarray}
\dot{u_i} &=& u_{i-1}\left(u_{i+1} -u_{i-2}\right)-u_i+F_m \label{l96m}\\
\dot{v_i} &=& v_{i-1}\left(v_{i+1} -v_{i-2}\right)-v_i+F_s+\gamma(u_i-v_i)
\label{l96s}
\end{eqnarray}
with $i=1,\ldots,L=1024$, and periodic boundary conditions: $u_0=u_L$,
$u_{-1}=u_{L-1}$, and $u_{L+1}=u_1$. The external forcing is controlled by
parameter $F_{m,s}$ and in the spirit of Lorenz formulation of the model it
represents the input of energy coming from sun radiation. We chose $F_m=8$, and
$F_s=F_m+\Delta F$ for the master and slave systems, respectively. (We have also
considered a mismatch in the dissipative term finding no significant
difference.) With this value of the external forcing (or larger) the system
exhibits (extensive) spatio-temporal chaos~\cite{pazo08,karimi10}. The numerical
integration was carried out using a fourth-order Runge-Kutta algorithm with time
step ${\mathrm d}t=0.01$. We study synchronization error as the parameter
mismatch $\Delta F$ is varied.

\subsubsection{Time-delay system}
\label{delay}
The third model we consider consists of two coupled time-delayed ordinary
differential equations:
\begin{eqnarray}
\dot u &=& -a u + b_m R(u_\tau) \label{mgm}\\
\dot v &=& -a v + b_s R(v_\tau) +\gamma (u-v) \label{mgs}
\end{eqnarray}
where $u_\tau\equiv u(t-\tau)$, likewise for $v$. Time-delay systems play a
fundamental role in the majority of studies on synchronization between
high-dimensional chaotic systems. After the change of variables~\cite{arecchi92}
\begin{equation}
 t=x+\vartheta\tau
\label{tx}
 \end{equation}
the delay system transforms into a one-dimensional spatially extended system
with the spatial coordinate $x\in[0,\tau)$, and evolving with a discrete
temporal variable $\vartheta=0,1,\ldots$. The perturbation dynamics of
time-delay systems, in the large ``size'' limit ($\tau\gg 1$), is in all
respects equivalent to that of standard spatio-temporal chaotic
systems~\cite{pik98,pazo10}. As a typical example we studied the Mackey-Glass
system~\cite{mg77}:
$$
R(\rho)=\frac{\rho}{1+\rho^{10}}.
$$
The model parameters are set to $a=1$, $b_m=2$, $b_s=b_m+\Delta b$. The chosen
delay time $\tau=2000$ is large enough to make the system highly chaotic.
Complete synchronization for identical systems, $b_m=b_s$, was
shown~\cite{szendro05} to belong to the bKPZ class, as occurs for standard
spatially extended chaotic systems. The numerical integration used was a
third-order Adams-Bashforth-Moulton predictor-corrector method~\cite{Press} with
time step $dt=0.02$ t.u.

\subsection{Spatially-extended system with unresolved scales}

In Sec.~\ref{unresolved} we shall consider the case of synchronization of an
imperfect slave system that does not contain the variables corresponding to the
small scale dynamics present in the master. This two-scale situation is
typically found in meteorology, for instance, where the models necessarily do
not incorporate small-scale turbulence present in the actual atmosphere. The
two-scale version of the Lorenz-96 model~\cite{lorenz96} is customarily used
to study this effect by supplementing Eq.~(\ref{l96m}) with a second
ring of fast evolving variables $y$. The complete system writes:
\begin{subequations}
\label{truth}
\begin{align}
& \dot u_i=u_{i-1}(u_{i+1}-u_{i-2})-u_i+F-
\frac{\eta \,c}{b}\sum_{j=1}^{J} y_{j,i}\label{e.eq2a}\\
& c^{-1} \dot y_{j,i}=b\, y_{j+1,i}(y_{j-1,i}-y_{j+2,i})-y_{j,i}+
\frac{|\eta|}{b} u_i,\label{e.eq2b}
\end{align}
\end{subequations}
where $i=1,\dots,L=36$, $j=1,\dots,J=10$, and periodic boundary conditions are
assumed: $u_{L+1}=u_1$, $u_{0}=u_L$, $u_{-1}=u_{L-1}$, and $y_{J+1,i}=y_{1,i+1}$,
$y_{J+2,i}=y_{2,i+1}$, $y_{0,i}=y_{J,i-1}$. The constant $c=10$ makes the
characteristic time scale of the $y$ variables to be 10 times shorter than
for the $u$ variables. The constant $\eta$ controls the coupling between
the two layers, while other constants are taken here as originally adopted by
Lorenz: $b=F=10$. With respect to the original formulation of the model, we have
replaced $\eta$ by $|\eta|$ in Eq.~(\ref{e.eq2b}) in order to consistently
explore also negative values of $\eta$. For $\eta=0$ the fast scale becomes
irrelevant and the system is fully described by Eq.~(\ref{e.eq2a}) alone. 

We adopt Eq.~(\ref{truth}) for the master, while the slave system (lacking 
the fast variables $y$) is governed by Eq.~(\ref{l96s}) with the same forcing
constant $F=10$ and size $L=36$ than the master. The numerical simulations were
carried out using a fourth-order Runge-Kutta algorithm with time step
$dt=10^{-3}$.

\section{Short overview of complete synchronization of spatio-temporal chaos}
\label{review}

In order to construct a stochastic field description of synchronization under
finite imperfections, we build upon the existing picture for synchronization
between identical systems
with smooth equations.
An interesting and far-reaching result~\cite{pik94,pik98,pazo08} is that, for
spatially extended systems, the spatio-temporal scaling properties of
infinitesimal perturbations $\delta u$ are generically captured by a linear
stochastic equation:
\begin{equation}
\partial_t \delta u =\xi \delta u +\partial_{xx} \delta u
\label{multiplicative0}
\end{equation}
where $\xi(x,t)$ is a white noise. Under a Hopf-Cole transformation
\begin{equation}
h(x,t)=\ln|\delta u(x,t)| ,
\end{equation}
and assuming the Stratonovich interpretation for the noise, we get the
Kardar-Parisi-Zhang (KPZ) equation~\cite{kpz},
\begin{equation}
 \partial_t h = \xi + (\partial_x h)^2 + \partial_{xx} h .
\label{kpzeq}
\end{equation}
It has been clearly demonstrated by Pikovsky {\it et al.}~\cite{pik94,pik98}
that the main Lyapunov vector obeys scaling laws in perfect agreement with the
universality class defined by KPZ, see also Paz\'o and L\'opez~\cite{pazo10}
for equivalent results for time-delayed systems.

Following the same approach, Ahlers and Pikovsky~\cite{ahlers02} proposed the
following stochastic field equation for the synchronization error
$w(x,t)=v(x,t)-u(x,t)$ between identical systems:
\begin{equation}
\partial_t w =(\xi-\gamma) w +\partial_{xx} w - p |w|^{p+1}
\label{multiplicative}
\end{equation}
($p=2$ was explicitly used by Ahlers and Pikovsky~\cite{ahlers02}). This is
basically the stochastic equation~(\ref{multiplicative0}) supplemented with a
nonlinear saturation term~\cite{tu97}. Equation~(\ref{multiplicative}) with
$p>0$ defines the multiplicative-noise (MN) universality
class~\cite{mamunoz98,mamunoz03b} (also denoted MN1 class for the sake of
distinguishing it from the complementary MN2 class corresponding to $p<0$).

Let $\gamma_c$ be the synchronization threshold such that the system fails to
synchronize for $\gamma<\gamma_c$, while complete synchronization ($w(x,t)=0$)
is the asymptotic state of Eq.~(\ref{multiplicative}) for $\gamma>\gamma_c$. For
smooth systems $\gamma_c$ is precisely the point at which the synchronized state
undergoes a transition from linear unstable to linearly stable.

\begin{figure}
\centerline{\includegraphics*[width=7cm,clip=true]{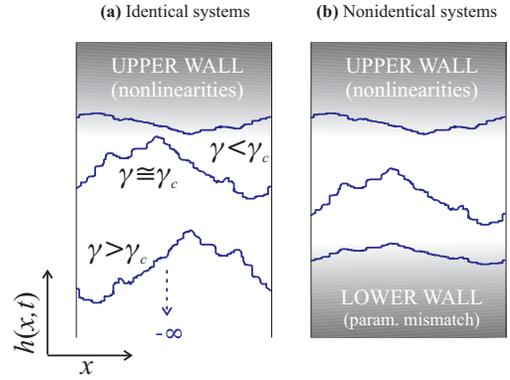}}
\caption{The curves sketch snapshots of $h(x,t)=\ln|w(x,t)|$ for different
couplings $\gamma$. (a) Identical systems: Above threshold ($\gamma>\gamma_c$)
the synchronization error drops to zero ($h\to-\infty$). As $\gamma$ decreases
from $\gamma_c$, $\bar h(t)$ increases but the width of $h(x,t)$ decreases. (b)
Non-identical systems: Parameter mismatches act as a lower wall that prevents
the synchronization error from dropping to zero even for large $\gamma$ values.
The maximum width of $h$ is roughly observed when $h$ is about equidistant
from both upper and lower walls.}
\label{esquema}
\end{figure}

An interesting observation is that under a Hopf-Cole transformation, $h=\ln|w|$,
Eq.~(\ref{multiplicative}) yields the so-called bounded
Kardar-Parisi-Zhang~\cite{tu97} (bKPZ) equation:
\begin{equation}
 \partial_t h = \xi-\gamma  + (\partial_x h)^2 + \partial_{xx} h - p e^{p h} .
\label{bkpz}
\end{equation}
The growth-limiting $e^{ph}$ term is a bound (or `upper wall') that prevents $h$
from going to infinity for $\gamma<\gamma_c$, while for $\gamma>\gamma_c$ one
has $h\to-\infty$ as time evolves (i.e.~the synchronization error vanishes).
Fig.~\ref{esquema}(a) sketches snapshots of the log-transformed error $h$ for
different values of $\gamma$. An important side effect of the upper wall is the
progressive reduction of the width of the ``surface'' $h$ as $\gamma$ decreases
from $\gamma_c$. 

The bKPZ Eq.~(\ref{bkpz}) was proposed~\cite{ahlers02} as the minimal model of
synchronization between extended systems, and this has been further confirmed
by Szendro {\it et al.}~\cite{szendro09} and Ginelli {\it et
al.}~\cite{ginelli09}, and Szendro and L\'opez~\cite{szendro05} for time-delay
systems. Though a more complicated field equation could in principle
occur~\cite{mamunoz03}, the bKPZ class (or its equivalent MN class) is
quantitatively consistent with the numerical results.

Parameter mismatch between master and slave must have an effect that necessarily
changes the picture  presented in Fig.~\ref{esquema}(a). Note that the absorbing
state at $h=-\infty$ does not exist anymore, since complete synchronization,
$w=0$, is no longer possible. The simplest way to implement the breakdown of the
synchronized state is to introduce a new repulsive lower wall such that the
error surface $h$ is kept away from $-\infty$, as sketched in
Fig.~\ref{esquema}(b). The joint effect of the upper and lower walls will be an
error surface that lies somewhere in between. The mathematical form of the lower
wall remains to be defined and will become clear after we discuss some numerical
results. Suffices to say now that the lower wall is well described by a term
proportional to the parameter mismatch, irrespective of the microscopic details
of the model.

\section{Numerical Results: Parameter mismatch}
\label{mismatch}

In this section we present the results of our simulations and give a rationale
behind our proposed picture in Fig.~\ref{esquema}(b) for non-identical systems.
We postpone to Sec.~\ref{concl} deeper considerations about the results.

\begin{figure}
\centerline{\includegraphics*[width=7cm,clip=true]{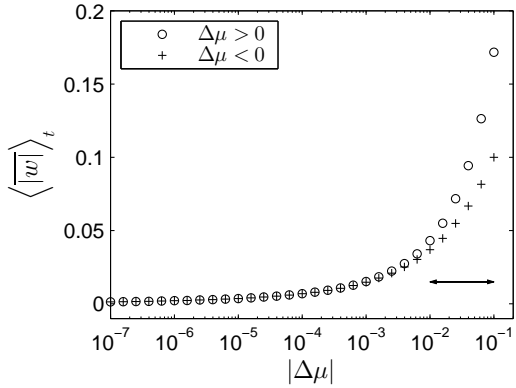}}
\caption{Average absolute-valued synchronization error at criticality
$\gamma=0.289376 \simeq \gamma_c$ for different values of $\Delta\mu$.
The double sided arrow indicates the values, finite and
non-infinitesimal, for the mismatch range
considered in this paper.}
\label{nonsmall}
\end{figure}

\subsection{Coupled-map lattice}

The error field $w(x,t)=v(x,t)-u(x,t)$ for coupled CMLs, see Eqs.~(\ref{cmlm})
and (\ref{cmls}), is governed by
\begin{equation}
w(x,t+1)=(1-\gamma)(1+\varepsilon {\cal D}) \left[f_s(u(x,t)+w(x,t)) 
-f_m(u(x,t)) \right]
\label{w_eq}
\end{equation}
Inserting the explicit form of $f_{m,s}$ we obtain
\begin{equation}
w(x,t+1)=(1-\gamma)(1+\varepsilon {\cal D}) \left[\Delta\mu - 
2 u(x,t) w(x,t) - w(x,t)^2 \right]
\label{w_exp}
\end{equation} 
where $\Delta\mu=\mu_s-\mu_m$ is the parameter mismatch. For $\Delta\mu=0$ the
synchronization threshold is at $\gamma_c \simeq 0.289376$. In
Fig.~\ref{nonsmall} we may see that as $|\Delta\mu|$ increases the
space-averaged absolute synchronization error 
 \begin{equation}
 \overline{|w|}(t)=\frac{1}{L} \sum_{i=1}^L |w(x,t)| 
 \end{equation}
grows accordingly. It is worth to stress here that parameter mismatch values
considered in this work correspond to the region marked by a double-headed arrow
in Fig.~\ref{nonsmall}, these values are large and away from critical scaling
region $|\Delta \mu| \to 0$. Also note the different curves for
positive and negative $\Delta\mu$. 

\begin{figure*}
\centerline{\includegraphics *[width=0.8\textwidth,clip=true]{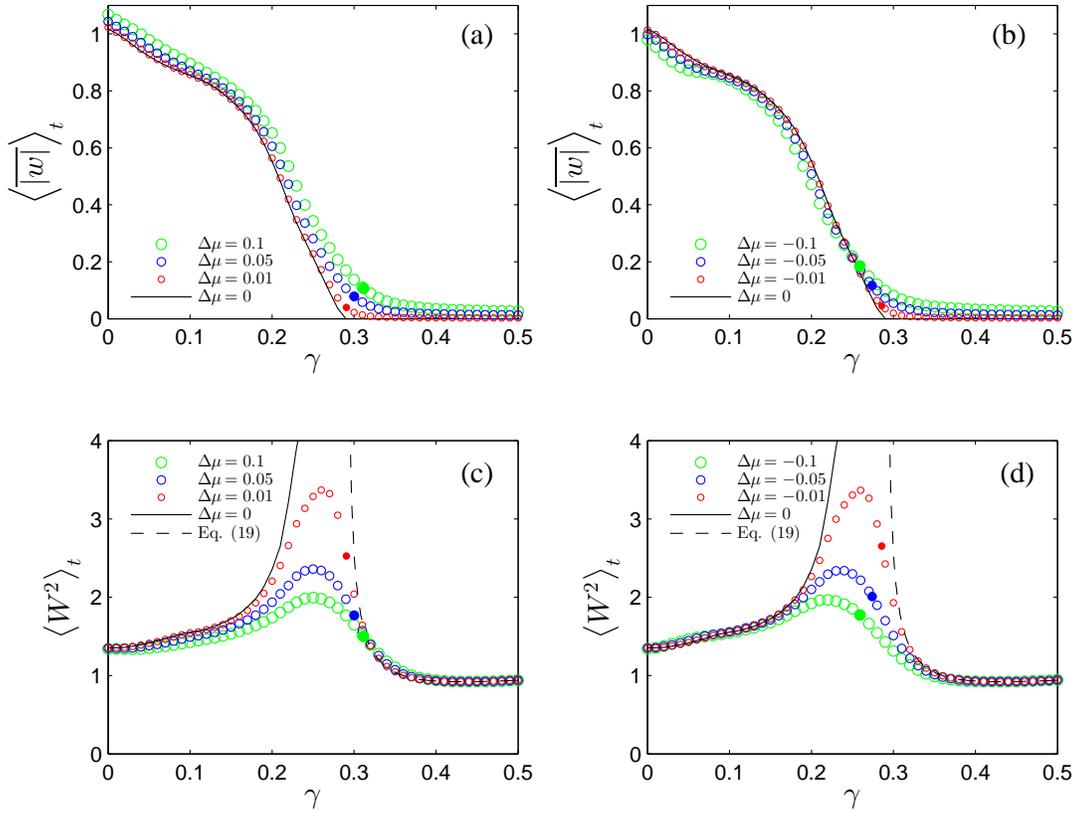}}
\caption{Coupled CMLs in Eqs.~(\ref{cmlm}) and (\ref{cmls}): (a,b) Average
absolute error as a function of $\gamma$ for different values of $\Delta\mu$.
(c,d) Time-averaged $W^2$ as a function of $\gamma$. Solid and dashed lines are
obtained from Eqs.~(\ref{uw}) and (\ref{lw}), respectively. Filled symbols
signal the points corresponding to the threshold of generalized synchronization,
$\gamma=\gamma_g$.}
\label{v_logistico}
\end{figure*}

In Fig.~\ref{v_logistico}(a,b) we plot the results of computing the temporal
average of the space-averaged error $\left< \overline{|w|}\right>_t$ for
different values of the mismatch $\Delta\mu$. For $\Delta\mu=0$ (solid line)
$|w|$ drops to zero at the complete-synchronization threshold $\gamma=\gamma_c$,
while for $\Delta\mu\ne0$ $\left< \overline{|w|}\right>_t$ decreases
monotonically as $\gamma$ increases but remains finite.

The statistic $\left< \overline{|w|}\right>_t$ 
does not inform about the homogeneity (or Gaussianity) of the synchronization
error across space.
This is conveniently analyzed
by measuring the 
width of the error surface $h=\ln|w|$, which is quantified by the
variance~\cite{szendro09}:
\begin{equation}
W^2=\frac{1}{L} \sum_{x=1}^L [h(x,t)-\bar h(t)]^2.
\end{equation}
This statistic quantifies of how many orders of magnitude the error amplitude
$|w|$ spans, {\it i.e.} the degree of error localization in space. Large values
are associated with a strong localization of the error at certain
sites~\cite{szendro09}, and a non-Gaussian distribution of $|w|$ across space
(``spatial intermittency'', so to speak). We may see in
Figs.~\ref{v_logistico}(c,d) that $\left< W^2\right>_t$ is non-monotonic and
reaches its maximum value at intermediate values of $\gamma$.

Equation (\ref{w_exp}) has two nonlinear terms, which 
suggests that we may consider two limits: 

\begin{enumerate}
\item[(i)] Large $|w|$ (small $\gamma$): in this limit we can neglect
the $\Delta\mu$ term recovering the equation 
\begin{eqnarray}
 w(x,t+1) = (1-\gamma)(1+\varepsilon {\cal D}) \times \nonumber \\
 \left[- 2 u(x,t) w(x,t) - w(x,t)^2 \right]
\label{uw}
\end{eqnarray}
for identical systems, where $|w|$ is known to be described by the MN
equation~(\ref{multiplicative}) or its variant the bKPZ equation~(\ref{bkpz}),
see also Fig.~\ref{esquema}(a).

\item[(ii)] Small $|w|$ (large $\gamma$): in this limit we can neglect the
quadratic term (and
any other higher-order term if they in fact exist) to get
\begin{eqnarray}
 w(x,t+1)=(1-\gamma)(1+\varepsilon {\cal D}) \times \nonumber\\
 \left[\Delta\mu - 2 u(x,t) w(x,t) \right] .
\label{lw}
\end{eqnarray}
It is convenient to stress that ---irrespective of the particular systems
involved--- the constant term $O(w^0)$ is a pure contribution of the parameter mismatch. In
some systems, the parameter mismatch might also yield small
irrelevant corrections in the linear $O(w^1)$ and/or higher-order terms
$O(w^2), O(w^3),\ldots$.
\end{enumerate}

In Fig.~\ref{v_logistico}(c,d) the solid and dashed lines are obtained
simulating Eqs.~(\ref{uw}) and~(\ref{lw}), respectively. The overall behavior of
$h=\ln|w|$, is like a KPZ surface in the presence of two bounds or walls: an
upper wall---already present for $\Delta\mu=0$---coming from quadratic (or
higher-order) terms, and a lower wall introduced by the parameter mismatch.
Figure~\ref{esquema}(b) sketches this simple (non-rigorous) picture: if
$\gamma<\gamma_c$ ($\gamma>\gamma_c$) the upper (lower) prevents the ``surface''
$h$ from diverging to $+\infty$ ($-\infty$). One may see  that, irrespective of
the sign of $\Delta\mu$, as $|\Delta\mu|$ decreases the point of maximal $\left<
W^2\right>_t$ moves to larger values of $\gamma$. This is consistent with our
interpretation of a lower wall that retreats to $-\infty$ as $|\Delta\mu|\to0$.
At $\gamma=0$ master and slave evolve independently and we obtain a value of
$\left< W^2 \right>_t$ consistent with the expectation for Gaussian
distributed error: $\left< W^2\right>_t=\pi^2/8=1.2337\ldots$; For large
$\gamma$ values the width becomes even smaller than this value due to the strong
pushing against the (lower) wall represented by the dashed line in
Fig.~\ref{v_logistico}(c,d).

\subsubsection*{Generalized synchronization} 

For the sake of comparison, 
a filled symbol for each data set is shown in
Figs.~\ref{v_logistico}, \ref{v}, and \ref{v_mg}. This point
marks the onset of generalized synchronization (GS), observed for $\gamma>\gamma_s$,
and determined by the `auxiliary system method'~\cite{abarbanel96}. In this
method an auxiliary system (also called replica), identical to the slave system,
is evolved in parallel with the same forcing from the master, and the convergence
of the slave and its replica defines the GS.
Although more stringent
definitions of GS exist~\cite{afraimovich86,rulkov95,kocarev96g}, 
the `auxiliary system method'~\cite{pecora97,parlitz97}
is intrinsically interesting because it signals the collapse of an ensemble of
slaves if such a setup were used (for instance in data assimilation
applications). The auxiliary system criterion has been often used in previous
works on GS of spatio-temporal chaos, which include an experimental realization
with a liquid crystal spatial light modulator with optoelectronic
feedback~\cite{rogers04}, a numerical integration of the complex Ginzburg-Landau
equation~\cite{hramov05}, and a reaction-diffusion system with a minimal
excitable dynamics~\cite{berg11}. There is also a good number of studies devoted
to GS in time-delayed systems~\cite{boccaletti00,zhan03,senthil07,senthil08}. 
The monotonic displacement of the GS threshold with the parameter mismatch, see Fig.~\ref{v_logistico},
stems from the changing chaoticity of the slave as its $\mu_s$ parameter is varied.
It is also interesting to note that the onset of GS evidences
an important difference between the upper wall and the lower wall.
Though both walls act similarly preventing the surface $h$ from reaching $\pm\infty$
while reducing the width,
the lower wall --in opposition to the upper wall-- does not induce
differences between the states of the replicas.

\subsection{Lorenz-96 model}

\begin{figure*}
\centerline{\includegraphics *[width=0.8\textwidth,clip=true]{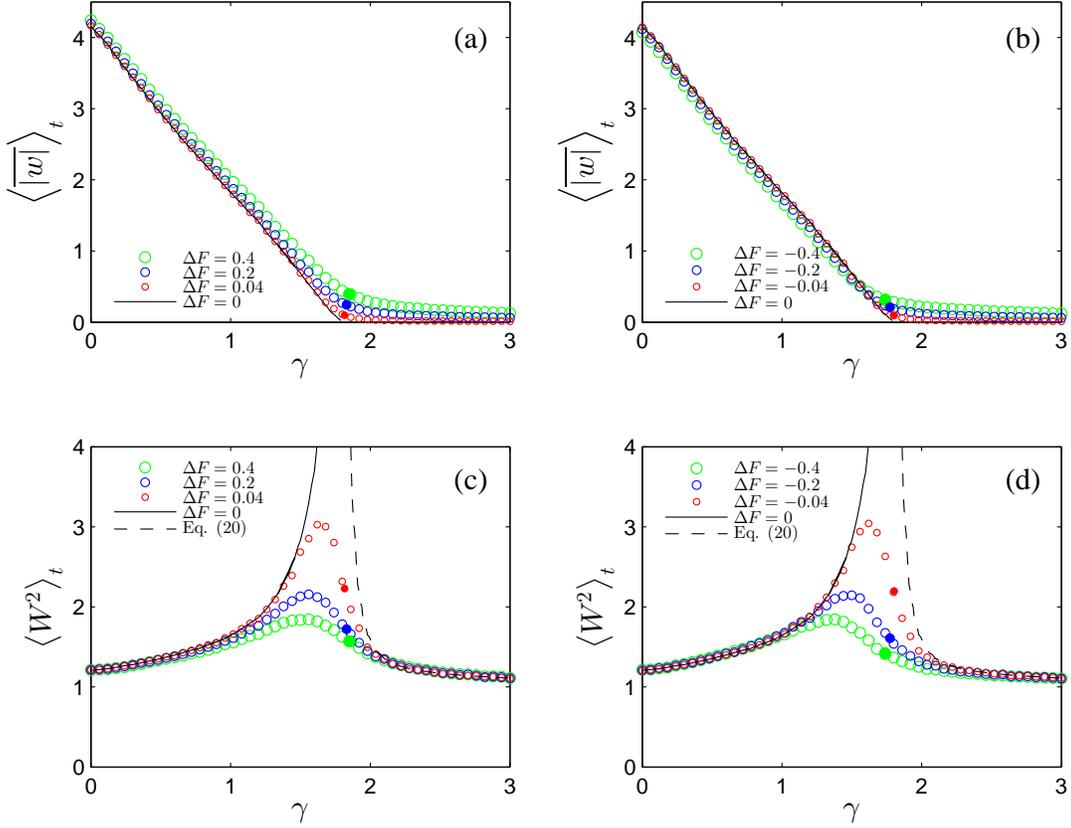}}
\caption{Coupled Lorenz-96 models in Eqs.~(\ref{l96m}) and~(\ref{l96s}):
(a,b) Average absolute error as a function of $\gamma$ for
different values of $\Delta F$. 
(c,d) Time-averaged $W^2$ as a function of $\gamma$. Filled symbols
signal the points corresponding to the threshold of generalized synchronization,
$\gamma=\gamma_g$.}
\label{v}
\end{figure*}

The results for the Lorenz-96 model, Eqs.~(\ref{l96m}) and~(\ref{l96s}),
are shown in Fig.~\ref{v}, and they are qualitatively identical to
those obtained in Fig.~\ref{v_logistico} for the CML. 
The surface width for large $\gamma$ can again be obtained 
by neglecting higher order terms $O(w^2)$ in the governing equations
for the error; analogously to Eq.~(\ref{lw}):
\begin{equation}
\dot{w_i} = \Delta F - (1+\gamma) w_i + w_{i-1}\left(u_{i+1} -u_{i-2}\right)+ u_{i-1}\left(w_{i+1} -w_{i-2}\right)
\end{equation}

\subsection{Time-delay system}
\label{sec_delay}

\begin{figure*}
\centerline{\includegraphics*[width=0.8\textwidth,clip=true]{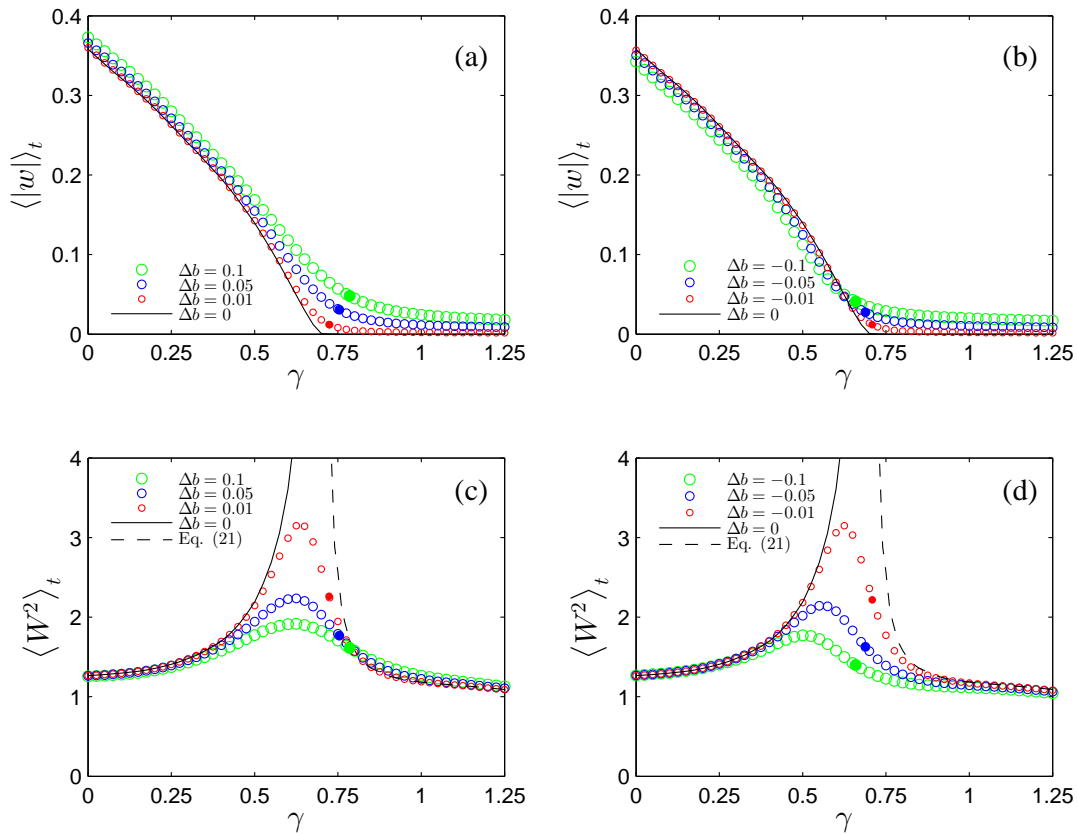}}
\caption{Coupled Mackey-Glass systems in Eqs.~(\ref{mgm}) and~(\ref{mgs}): (a,b)
Average absolute error as a function of $\gamma$ for different values of $\Delta
b$. (c,d) Time-averaged $W^2$ as a function of $\gamma$. Filled symbols
signal the points corresponding to the threshold of generalized synchronization,
$\gamma=\gamma_g$.}
\label{v_mg}
\end{figure*}

The results for the time-delayed systems, Eqs.~(\ref{mgm}) and~(\ref{mgs}), are
shown in Fig.~\ref{v_mg}. We have used the coordinate transformation~(\ref{tx})
and studied the system like a spatially extended system of length $\tau$. As
expected, we find no qualitative difference with respect to genuinely spatial
systems. Note that, when going back to the original temporal framework, $\left<
W^2\right>_t$ also serves to quantify the degree of intermittency (or
non-Gaussianity) of the signal $w(t)=v(t)-u(t)$. In the small error
approximation, neglecting higher order terms, we obtain:
\begin{equation}
 \dot w =\Delta b \, R(u_\tau) -(a+\gamma)w + b_m \,R'(u_\tau) \, w_\tau
\label{lw_mg}
\end{equation}

\section{Numerical Results: Unresolved scales}
\label{unresolved}

\begin{figure*}
\centerline{\includegraphics*[width=0.8\textwidth,clip=true]{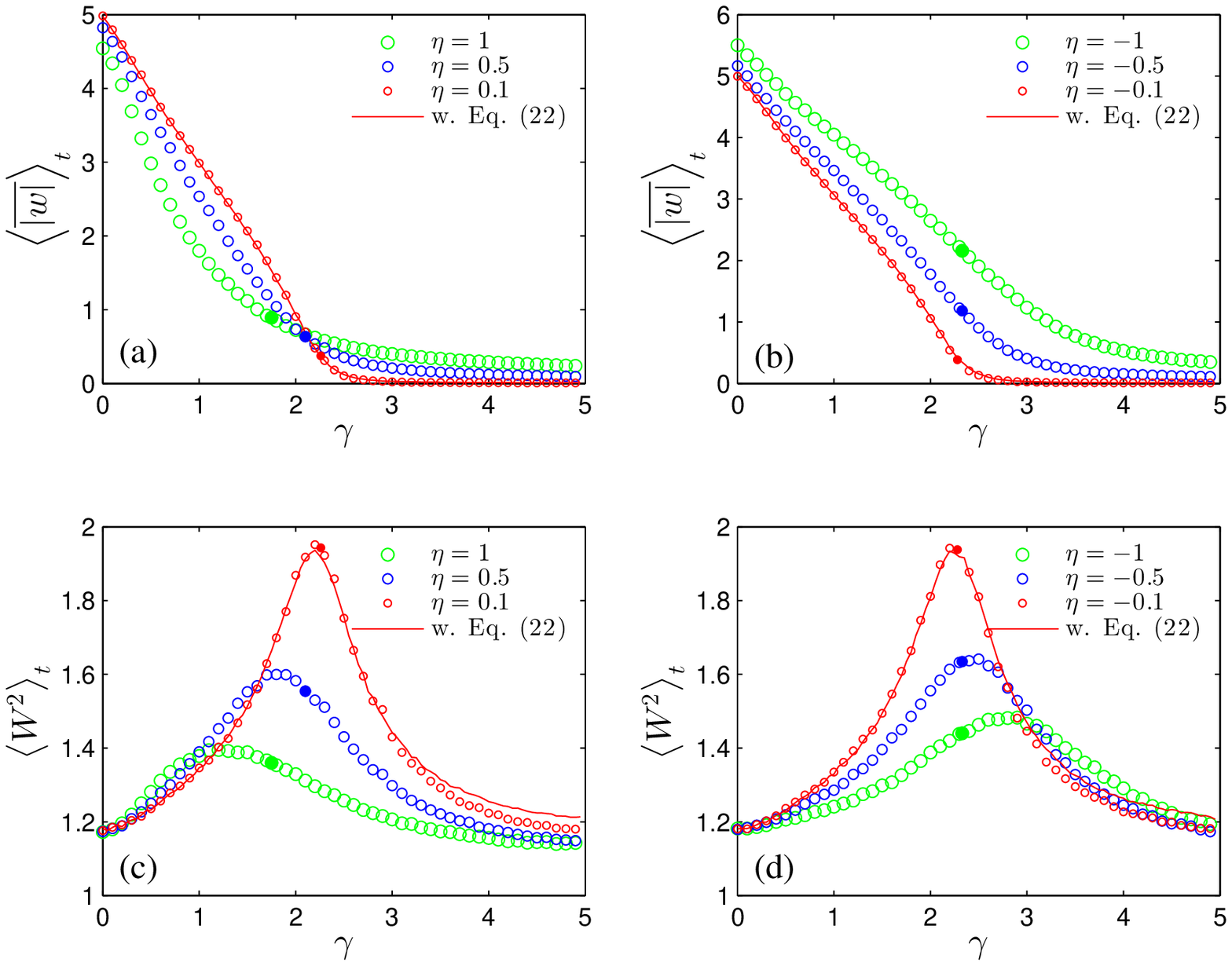}}
\caption{Two-scale Lorenz-96 model (master) coupled to a one-scale
Lorenz-96 model (slave).
(a,b) Average absolute error as a function of $\gamma$ for
different values of $\eta$. 
(c,d) Time-averaged $W^2$ as a function of $\gamma$.
In all panels the straight line is obtained when the master is replaced by
an approximation given in Eq.~(\ref{l96_adiabatic}) with $\eta=\pm0.1$.}
\label{l96-2}
\end{figure*}

The question of the limits to synchronization of two systems in a master/slave
configuration when the slave has a lower spatial resolution than the master is
very relevant in practical applications, including forecasting and data
assimilation in geoscience. The results with a two-scale Lorenz-96 model,
Eq.~(\ref{l96m}) for the master and Eq.~(\ref{truth}) as slave, are presented in
Fig.~\ref{l96-2}. It is remarkable that the two systems are able
to partially synchronize despite the obvious constitutive differences.

Interestingly, the statistical and dynamical properties of the synchronization
error are qualitatively similar to those obtained in the case of parameter
mismatch. Indeed, for very small $|\eta|$ one can see that the unresolved scales
lead to an effective mismatch since the fast variables $y$
follow almost immediately the $u$ variable: $y_{j,i}\simeq \frac{|\eta|}{b}u_i$. 
Hence
the $u$ variables of the master approximately obey the same one-scale equation
than the slave but with a mismatch in the dissipative term:
\begin{equation}
\dot{u}_i = u_{i-1}\left(u_{i+1} -u_{i-2}\right)-\left(1+\frac{\eta|\eta| J c}{b^2}\right) u_i+F \label{l96_adiabatic}\\
\end{equation}
Coupling the slave system to this approximation of the master for $\eta=\pm0.1$ we obtain
the solid lines in Fig.~\ref{l96-2}, finding a very good agreement with 
results obtained using the original master system in Eq.~(\ref{truth}).

\section{Discussion and Conclusions}
\label{concl}
All the numerical results presented so far can be understood as the combined
effect of an upper and lower wall. In particular, the dependence of $W^2$ on
$\gamma$ can be phenomenologically described by adding a term proportional to
the mismatch $\Delta \mu$ to Eq.~(\ref{multiplicative}):
\begin{equation}
\partial_t w =\Delta \mu \, \zeta+ (\xi-\gamma) w +\partial_{xx} w - p |w|^{p+1}
\label{multiplicative1}
\end{equation}
where $\zeta$ is a system-dependent (possibly fluctuating) term, whose detailed
form does not affect the argument. Basically, the same equation was proposed by
Ginelli {\it et al.}~\cite{ginelli09} to describe the scaling behavior at
$\gamma_c$ of $\left< \overline{|w|}\right>_t$ as $\Delta\mu \to 0$. In this
paper we are well above the scaling regime but nevertheless
Eq.~(\ref{multiplicative1}) can be invoked to understand the basic features
observed in Figs.~\ref{v_logistico}, \ref{v}, \ref{v_mg}, and \ref{l96-2}. This complements
the picture sketched in Fig.~\ref{esquema}.

At small $\gamma$ the error $|w|$ is large and the mismatch term in
Eq.~(\ref{multiplicative1}) becomes negligible so that the curves $\left<
W^2\right>_t$ converge for different values of the mismatch $\Delta\mu$, as
observed in Figs.~\ref{v_logistico}, \ref{v}, and \ref{v_mg}.  
In contrast, for large $\gamma$ the error $|w|$ becomes small, so the
$|w|^{p+1}$ term can be
neglected and we are led, therefore, to the equation:
\begin{equation}
\partial_t w =\Delta \mu \, \zeta+ (\xi-\gamma) w +\partial_{xx} w 
\label{multiplicative2}
\end{equation}
We can divide the equation by $\Delta \mu$ and 
define the scaled variable $w' = w/\Delta\mu$, which leads to the scaling
relation
\begin{equation}
\left< \overline{|w|}\right>_t \propto |\Delta\mu| 
\end{equation}
in agreement with our simulations (not shown), and Eqs.~(\ref{lw})-(\ref{lw_mg}). Moreover, note that
Eq.~(\ref{multiplicative2}) preserves the value of $W^2$ under changes of
$\Delta\mu$ since it can be absorbed by a shift of the surface position
$h'=h-\ln\Delta\mu$. This explains the collapse of the curves $\left< W^2
\right>_t$ at large $\gamma$ values for different intensities of the mismatch.

A final remark on the nature of the lower wall is in order. We may see that a
straightforward application of the Hopf-Cole transformation to
Eq.~(\ref{multiplicative2}) yields a KPZ equation with an extra term of the form
$\Delta\mu \, \zeta e^{-h}$, which is precisely a bKPZ equation with a
lower-wall~\cite{mamunoz98}. However, the exact correspondence with a bKPZ-type
equation does not hold at least for two (not completely independent) reasons.
Firstly, the function $\zeta$ is expected to be a fluctuating quantity in
general, which is known~\cite{mamunoz04b} to have an important effect on
assigning the corresponding universality class. Secondly, the absolute value
$|w|$, which is implicit in the Hopf-Cole transformation, cannot be simply
disregarded. In chaotic systems the sign of $w$ always fluctuates and the
existence of zeros of $w$ is an unavoidable
characteristic~\cite{szendro07,pazo08}. It turns out that these zeros do yield
divergent contributions to the suggested (sign-independent) order parameter
$\overline{|w|^{-1}}$ (see Al Hammal {\it et al.}~\cite{alhammal05} for further
details on this rather technical question). We stress, nonetheless, that even if
the lower wall is in mathematical terms an exponential bound with a
fluctuating amplitude it has effectively the same qualitative effect as a simple
exponential wall. 

In this work we have put the focus on the nontrivial statistical features of the
synchronization error between non-identical spatio-temporally chaotic systems.
In the framework of a stochastic field theory that describes the synchronization
error $w(x,t)$, the effect of the parameter mismatch is to give rise to a new
lower wall (see Fig.~1), so that the absorbing state $w =0$ cannot be reached.
The qualitative behavior of the `error surface' $h=\ln|w|$  can be thus
understood in terms of the competition between the two opposing walls: the
standard upper wall that appears due to nonlinear saturation for large $|w|$,
and the new lower wall associated with the model mismatch. The combined effect
of the two walls permits the synchronization error to exhibit a significant
deviation from Gaussianity only in a certain coupling range about the
generalized synchronization threshold. Although our results have been obtained
for systems with unidirectional coupling (master/slave configuration), they
should apply equally well to the case of bidirectional coupling. The extension
of our results to nonsmooth systems (generically associated with the directed
percolation class~\cite{grassberger99,ahlers02,szendro05,ginelli09}) remains as
a task for future investigations.


We thank Alberto Carrassi for interesting discussions.
DP~acknowledges support by Cantabria International Campus, and by
Ministerio de Econom\'ia y Competitividad (Spain)
under the Ram\'on y Cajal programme.
%


\end{document}